\documentclass[twocolumn,           
               showpacs,            
               nopreprintnumbers,   
               aps,                 
               prd,          	    
               a4paper,             
               groupedaddress,      
               unsortedaddress,
               nofootinbib,         
               tightenlines,        
               floats               
               ]{revtex4}

\usepackage{graphicx}
\usepackage{amssymb,latexsym}
\usepackage[draft=false]{hyperref}

\begin{document}




\title{Nflation: non-gaussianity in the horizon-crossing approximation}
\author{Soo A Kim and Andrew R.~Liddle}
\affiliation{Astronomy Centre, University of Sussex, 
             Brighton BN1 9QH, United Kingdom}
\date{\today} 
\pacs{98.80.Cq \hfill astro-ph/0608186}
\preprint{astro-ph/0608186}


\begin{abstract}
We analyze the cosmic non-gaussianity produced in inflation models
with multiple uncoupled fields with monomial potentials, such as
Nflation. Using the horizon-crossing approximation to compute the
non-gaussianity, we show that when each field has the same form of
potential, the prediction is {\em independent} the number of fields,
their initial conditions, and the spectrum of masses/couplings. It
depends only on the number of $e$-foldings after the horizon crossing
of observable perturbations. We also provide a further generalization
to the case where the fields can have monomial potentials with
different powers.  Unless the horizon-crossing approximation is
substantially violated, the predicted non-gaussianity is too small to
ever be observed.
\end{abstract}

\maketitle

\section{Introduction}

There has been recent interest in models of inflation with multiple
uncoupled fields, an example being the Nflation model of 
Dimopoulos et al.~\cite{DKMW} which corresponds to a collection of
massive fields. Inflation may proceed more efficiently in such
scenarios due to the assisted inflation phenomenon \cite{LMS}, and the
models may be well-motivated within the context of string theory
or dimensional reduction \cite{KO,mfield,DKMW,EM}.

Various observational predictions have been made for such
scenarios. Alabidi and Lyth \cite{AL} made a comprehensive study of
the case of massive fields, demonstrating that the tensor-to-scalar
ratio $r$ always takes the single-field value $r=8/N$, where $N$ is
the number of $e$-foldings since horizon crossing, independently of
the mass spectrum and the initial conditions. This result was extended
to monomial potentials by Piao \cite{piao}. Alabidi and Lyth also
showed that the spectral index $n_{{\rm s}}$ was model dependent, but
always less than the single-field value, as previously shown for two
massive fields by Lyth and Riotto \cite{LR}. Actual values of $n_{{\rm
s}}$ were evaluated for a particular choice of mass spectrum and
initial conditions by Easther and McAllister \cite{EM}, and for
various choices of the number of fields $N_{{\rm f}}$ and mass
spectrum with random field initial conditions by Kim and Liddle
\cite{KL}.

A key interest of multi-field models is whether they can generate
significant non-gaussianity \cite{Mald,LRod,SL,AL,RST,VW}. The emerging
view is that the non-Gaussianity is always small in models of the type
considered here (though see Ref.~\cite{RST}). Alabidi and Lyth
\cite{AL} computed the non-linearity parameter $f_{{\rm NL}}$ using
the separate Universes approach, obtaining a formula claimed to
indicate that it is always less than unity though it was not
explicitly calculated for any models. Vernizzi and Wands \cite{VW} did
explicitly evaluate a similar formula for the case of two massive
uncoupled fields, indicating that it is indeed suppressed by the
values of slow-roll parameters at horizon exit and hence much less
than unity.

In this article, we explicitly calculate the non-linearity parameter
$f_{{\rm NL}}$ for multiple uncoupled fields with monomial potentials,
i.e.
\begin{equation}
V = \sum_{i=1}^{N_{{\rm f}}} \lambda_i \phi_i^\alpha
\end{equation}
where $\alpha$ is an even positive integer (the same for each field)
and each of the $N_{{\rm f}}$ fields may have a different
mass/coupling $\lambda_i$. We use a simplified version of the
formalism of Vernizzi and Wands \cite{VW} by adopting the
horizon-crossing approximation, which in essence assumes that the
field trajectory becomes straight by the end of inflation or soon
after, and that isocurvature perturbations do not play a role
subsequently. Our work extends that of Alabidi and Lyth \cite{AL} by
explicitly evaluating the non-gaussianity expression for these models,
extends that of Vernizzi and Wands \cite{VW} by considering more than
two fields, and extends both by evaluating the result for general
(even) monomial potentials. We end by further generalizing to allow
each potential to have a different power-law index $\alpha_i$.

\section{The calculation}

We follow the notation of our earlier paper \cite{KL} and of Vernizzi
and Wands \cite{VW}. The calculation is a straightforward
implementation of those already in the literature. For a set of
uncoupled fields, the equation for the number of $e$-foldings $N$, in
the slow-roll approximation, is \cite{LR}
\begin{equation}
N \simeq  -\frac{1}{M_{{\rm Pl}}^2} \sum_i
\int_{\phi_i}^{\phi_i^{{\rm end}}} \frac{V_i}{V'_i} \, d\phi_i  \simeq 
 \frac{\sum_i \phi_i^2}{2\alpha M_{{\rm Pl}}^2} \,,
\label{e:efolds} 
\end{equation}
and the tensor-to-scalar ratio is \cite{piao}
\begin{equation}
\label{e:ts}
r  \simeq  \frac{8M_{{\rm Pl}}^2}{\sum_i (V_i/V'_i)^2} \simeq
\frac{4\alpha}{N}\,. 
\end{equation}
Here $V_i$ is the potential of the $i$-th field $\phi_i$, $V'_i \equiv
dV_i/d\phi_i$, $M_{{\rm Pl}}$ is the reduced Planck mass, and
throughout there are no summations unless indicated explicitly.  In
the last expression for $N$ the lower limits of the integrals,
corresponding to the end of inflation, can be neglected and have been.

An expression for the non-gaussianity can be obtained using the
separate Universes/$\delta N$ formalism \cite{Star,SS,WMLL,LRod}. The
non-linearity parameter $f_{{\rm NL}}$ is then given by
\cite{Mald,SL,VW} 
\begin{equation}
\label{e:vw}
-\frac{6}{5} \, f_{{\rm NL}} = \frac{r}{16}(1+f) + \frac{\sum_{i,j}
N_{,i} N_{,j} N_{,ij}}{\left( \sum_k N^2_{,k} \right)^2} \,,
\end{equation}
where `,$i$' indicates derivative with respect to $\phi_i$.  Here $r$
is the tensor-to-scalar ratio, given by Eq.~(\ref{e:ts}) for the
models we are discussing, and $f$ is a geometric factor relating to
the triangular bispectrum configuration being studied, lying in the
range $0 \leq f \leq 5/6$ \cite{Mald}. The first term is thus
guaranteed to be small by current observational limits on $r$
\cite{wmap3}. The second term is denoted $f^{(4)}_{{\rm NL}}$ and
needs to be computed.

We evaluate the second term using the horizon-crossing
approximation. This assumes that there will be a negligible correction
when shifting from an initially spatially-flat hypersurface to a final
uniform-density hypersurface. This is guaranteed if the trajectory
becomes straight before inflation ends (or even somewhat after), which
in multi-field models of the type we are studying should be typical
but cannot be absolutely generic.

In the two-field case, this was recently studied in detail by Vernizzi
and Wands \cite{VW}, who track the evolution of the perturbations
during inflation. Their expression for the non-gaussianity mostly
features terms evaluated at horizon crossing, plus one additional term
denoted $Z_c$. This term accounts for the contribution to the change
in $e$-foldings at the final uniform-density hypersurface, and evolves
during inflation driving evolution of $f_{{\rm NL}}$. If $Z_c$ is set
to zero, the formula Eq.~(\ref{e:fnl}) we give below is recovered. We
have reproduced their calculation, and find that while $Z_c$ is
substantial at horizon crossing in the specific case they analyze, it
becomes negligible by the end of inflation. Accordingly, our
expression is an excellent approximation to the desired answer, being
the one at the end of inflation, even though it is entirely evaluated
at horizon crossing. We expect the horizon-crossing approximation to
hold very well in typical situations (as already commented in
Ref.~\cite{AL}), though a more detailed analysis of this point is in
progress.

Using the horizon-crossing approximation, the derivatives of the
number of $e$-foldings can be written in terms of the potential as
$M_{{\rm Pl}}^2 \,N_{,i}\simeq V_i/V'_i$ \cite{LR}, leading to
\begin{eqnarray}
\label{e:fnl}
-\frac{6}{5} \, f^{(4)}_{{\rm NL}} & \simeq &  M_{{\rm Pl}}^2 
\left(\sum_j \frac{V_j^2}{V'^2_j}\right)^{-2}
\sum_i \frac{V_i^2}{V'^2_i} 
\left(1-\frac{V_i V''_i}{V'^2_i}\right) \quad \\
\label{e:fnlresult} & \simeq & \frac{\alpha M_{{\rm Pl}}^2}{\sum_i
\phi_i^2} \,.
\end{eqnarray}
Using Eqs.~(\ref{e:efolds}) and (\ref{e:ts}) immediately yields a
final answer
\begin{equation}
\label{e:result}
- \frac{6}{5} f_{{\rm NL}} = \frac{1}{2N} \left(2+f \right) =
  \frac{r}{8\alpha} \left( 2+f \right) \,,
\end{equation}
which is the main result of this paper. 

Equation (\ref{e:result}) matches exactly the result found by Vernizzi
and Wands \cite{VW}, but their result was calculated only for two
massive fields. We have shown that the same result holds for arbitrary
numbers of fields and for general (even) monomial potentials. Such
models are therefore highly predictive in their non-gaussianity, but
sadly the prediction is for a number so small that it is swamped by
effects of nonlinear gravity and can never be detected.

In fact we can even generalize this calculation further, by allowing
each field to have a different exponent $\alpha_i$:
\begin{equation}
V=\sum_{i=1}^{N_f} \lambda_i \phi_i^{\alpha_i}\,.
\end{equation}
Using Eqs.~(\ref{e:efolds}) and (\ref{e:ts}), the number of
$e$-foldings and the tensor-to-scalar ratio will be
\begin{eqnarray}
N & \simeq & \frac{1}{2M_{{\rm Pl}}^2}\sum_i
\frac{\phi_i^2}{\alpha_i}\,,\\ 
r & \simeq & \frac{8M_{{\rm Pl}}^2}{\sum_i (\phi_i/\alpha_i)^2}\,.
\end{eqnarray}
The first term of Eq.~(\ref{e:vw}) is unchanged, but the second now reads
\begin{eqnarray}
- \frac{6}{5} f^{(4)}_{{\rm NL}} & \simeq &  M_{{\rm Pl}}^2 
\frac{\sum_i \phi_i^2 / \alpha_i^3}
{\left(\sum_j \phi_j^2 / \alpha_j^2\right)^2}\,,\\
& \simeq & \frac{r^2}{64M_{{\rm Pl}}^2}\sum_i
\frac{\phi_i^2}{\alpha_i^3}\,,\\ 
& \simeq & \frac{1}{2N} \left[\frac{\sum_i \phi_i^2 / \alpha_i^3}
{\left(\sum_k \phi_k^2 / \alpha_k^2\right)^2} 
\sum_j \left( \frac{\phi_j^2}{\alpha_j}\right)\right]\,,
\end{eqnarray}
where we have written it in various equivalent forms. If the
$\alpha_i$ are all the same we recover the previous result
Eq.~(\ref{e:fnlresult}). However if the $\alpha_i$ are different the
result does depend on initial conditions and on the model parameters,
while still being slow-roll suppressed. The easiest way to see this is
to bear in mind that $\alpha_i \geq 2$, and use the second of the
above equations to obtain $|(6/5)f^{(4)}_{{\rm NL}}| \leq r/16$.

\section{Conclusions}

We have computed the non-linear parameter $f_{{\rm NL}}$ that measures
primordial non-gaussianity for models with multiple uncoupled fields,
generalizing calculations in the literature. We focussed mainly on the
case where the fields have monomial potentials with the same slope but
different amplitudes, e.g.~a set of massive fields with an arbitrary
mass spectrum. We have shown that within the horizon-crossing
approximation these models make a unique prediction for the amplitude
of non-gaussianity, independent of the field initial conditions, of
the number of fields, and of their mass/coupling spectrum. The
predicted non-gaussianity is however too small be measured. We also
generalized this result further to allow different power-laws for each
field.

Our calculation gives the perturbations associated with the
horizon-crossing epoch.  There is also the question of whether further
perturbations might be generated after inflation, for instance by a
curvaton-like mechanism (see e.g.~the discussion in
Ref.~\cite{VW}). Such effects would be absent if the late stages of
inflation are driven by a single field, but otherwise would depend on
the routes by which the scalar fields decay into conventional and dark
matter. We have assumed such effects are absent.

\begin{acknowledgments}
S.A.K.\ was supported by the Korean government and A.R.L.\ by PPARC
(UK). We thank Nicola Bartolo, Richard Easther, Antonio de Felice,
David Lyth, Karim Malik, Filippo Vernizzi, and David Wands for useful
discussions.
\end{acknowledgments}

\end{document}